\documentclass[12pt,draftcls,onecolumn]{IEEEtran}
\usepackage{graphicx}
\usepackage{bm}
\usepackage{url}
\usepackage{amsfonts,amssymb,amsmath}
\usepackage{multicol,multirow}
\usepackage[english]{babel}
\usepackage{tabularx}
\usepackage{float}
\usepackage{nomencl}
\usepackage{cite}

\makeatletter

\newcommand{\Rmnum}[1]{\expandafter\@slowromancap\romannumeral #1@}
\makeatother
\floatstyle{ruled}
\newfloat{program}{thp}{lop}
\floatname{program}{Algorithm}

\title{Adaptive Interference Alignment with CSI Uncertainty}
\author{
Baile Xie, {\em Student Member,~IEEE,}
    Yang Li, {\em Student Member,~IEEE,}\\
    Hlaing Minn, {\em Senior Member,~IEEE,} and Aria
    Nosratinia, {\em Fellow,~IEEE}

\thanks{Manuscript received September 02, 2011, revised Jul 03, 2012; accepted September 06, 2012. The associate editor coordinating the review of this paper
and approving it for publication was Xi Zhang.}
\thanks{B. Xie, Y. Li, H. Minn and A. Nosratinia are with the Department of Electrical Engineering, the University of Texas at Dallas, Richardson, TX, 75080, USA, e-mail: \{baile.xie, yang, hlaing.minn, aria\}@utdallas.edu}
\thanks{The work of A. Nosratinia is supported in part by the NSF under grant CCF1219065.}
}
\begin{document}
\maketitle

\begin{abstract}
Interference alignment (IA) is known to significantly increase
sum-throughput at high SNR in the presence of multiple interfering
nodes, however, the reliability of IA is little known, which is the
subject of this paper. We study the error performance of IA and
compare it with conventional orthogonal transmission schemes. Since
most IA algorithms require extensive channel state information (CSI),
we also investigate the impact of CSI imperfection (uncertainty) on
the error performance. Our results show that under identical rates, IA
attains a better error performance than the orthogonal scheme for practical signal to noise ratio (SNR) values but is
more sensitive to CSI uncertainty. We design bit loading algorithms
that significantly improve error performance of the existing IA
schemes. Furthermore, we propose an adaptive transmission scheme that
not only considerably reduces error probability, but also produces
robustness to CSI uncertainty.
\end{abstract}

\IEEEpeerreviewmaketitle


\section{Introduction}
Interference alignment (IA) achieves, at high signal-to-noise-ratio
(SNR), a sum-rate increasing linearly with the number of user pairs in
an interference network~\cite{CJafar2008}. The main idea of IA is to
coordinate transmitted signals so that the interferences are
concentrated in certain subspaces at the unintended receivers. This
opens up an interference-free subspace for each pair. IA achieves the
maximum multiplexing gain or degree of freedom (DoF) which is
significantly larger than conventional orthogonal schemes (e.g., time
division multiple access (TDMA)) or certain non-orthogonal schemes
(e.g., treating interference as noise)~\cite{CJafar2008}. This result
has inspired a great deal of research
activity~\cite{YJafar2010,GJafar2008,YSung2010,SPRH2009AlterMin,Thukral2009,
  Varanasi2010,Baile2011Globecom,DownlinkIA2011,CellularBlindIA2011ICC,IADiverisity2011,IASINR2011}.

Many of the existing works study IA from a capacity (multiplexing
gain)
perspective~\cite{CJafar2008,GJafar2008,YJafar2010,YSung2010,SPRH2009AlterMin,Thukral2009,
  Varanasi2010,Baile2011Globecom,
  DownlinkIA2011,CellularBlindIA2011ICC}, but very few works
analyze the reliability and/or error performance of IA. 
Motivated by its practical importance, this work investigates the
error performance of IA and compare it with conventional methods
(e.g. TDMA with singular value decomposition (SVD)-based spatial
multiplexing (SM)). In addition, most IA schemes require extensive
channel state information (CSI), which is imperfect in practice. This
paper also studies the impact of CSI imperfection (uncertainty/error)
on IA. We propose an adaptive method as well as bit loading algorithms for IA
schemes, which produce considerable gains over existing IA
methods. Specifically,
\begin{itemize}
\item
We analyze two representative IA algorithms: minimizing interference
leakage algorithm (MinIL) and maximizing signal to noise plus
interference ratio (SINR) algorithm (Max-SINR)~\cite{GJafar2008}.  We
find that under perfect CSI, although both algorithms achieve the same
multiplexing gain, Max-SINR attains much lower bit error rate (BER)
than MinIL (more than 7 dB SNR gain), and it always outperforms the
conventional SVD-based SM.  In contrast, MinIL has higher BER than
SVD-based SM for some network configurations\footnote{The network
  configuration refers to the number of user pairs, the numbers of
  transmit antennas and receive antennas in an interference network.},
even though it achieves larger multiplexing gain. We also obtain a
closed-form BER expression for the MinIL algorithm.

\item
Our results show that among the compared schemes, Max-SINR is the most
sensitive to CSI uncertainty, followed by MinIL, and SVD-based SM is
the least sensitive to imperfect CSI. Nevertheless, Max-SINR still
achieves the smallest BER if CSI uncertainty is less than 10\% (at SNR
20 dB); if the CSI uncertainty exceeds 10\%, all the three schemes
have almost identical (poor) BER performances. Moreover, we find that IA
algorithms exhibit error floors when CSI is imperfect.

\item
We propose IA bit-loading algorithms which significantly reduce BER for
both MinIL and Max-SINR, producing 6 dB and 4 dB SNR gain (at BER of
$10^{-2}$) respectively.

\item
We devise an adaptive transmission scheme that switches among the
three schemes. Adaptive transmission achieves 5 dB SNR gain
compared with the best of the three modes (with bit loading), and is
also more robust to CSI uncertainty than Max-SINR.
\end{itemize}

Some of the related literature on IA is as follows. In the presence of
perfect CSI, the feasibility condition of IA in
multiple-input-multiple-output (MIMO) interference networks with
constant channel coefficients was investigated
in~\cite{YJafar2010}. In~\cite{GJafar2008,YSung2010,SPRH2009AlterMin},
different algorithms were proposed to design the precoding and receive
combining vectors in IA. Recently, IA with partial CSI has attracted
significant attention. For single-input-single-output (SISO) network,
B\"{o}lcskei and Thukral~\cite{Thukral2009} studied the achievable
multiplexing gain of IA when CSI was obtained via limited feedback,
which was extended to MIMO network by Krishnamachari and
Varanasi~\cite{Varanasi2010}. Xie~et al.~\cite{Baile2011Globecom}
found the optimal numbers of feedback bits and cooperative user pairs
so that the overall throughput was maximized at high SNR. Most
recently, Ning et al. \cite{IADiverisity2011} investigated the
reliability issue of IA from diversity perspective, showing conditions
for achievability of diversity gain in IA. For the IA zero-forcing
algorithm, Nosrat-Makouei et al.~\cite{IASINR2011} found approximate SINR expressions in the presence of imperfect CSI and
channel correlation.

The rest of the paper is organized as follows: Section \ref{SecSigModl}
introduces the signal model and the background of IA. Section
\ref{SecBER} discusses transmission schemes for the interference
channel and their corresponding BER analyses. In Section \ref{SecBL},
the bit loading algorithm and the adaptive transmission scheme are
presented. Then, we assess the effects of CSI uncertainty on BER
performances in Section \ref{SecCSI}. Section \ref{SecSimulation}
presents the simulation results and corresponding discussions, and
Section \ref{SecCon} concludes this paper.

\emph{Notation}: Throughout the paper, boldface lower-case letters
 stand for vectors while upper-case letters
represent matrices.  $\textbf{A}^{\dag}$ indicates the Hermitian
transpose of $\textbf{A}$.  $\left \|\textbf{a}\right \| $ means
$\ell_2$-norm. $\mathcal{CN}(\textbf{a},\textbf{A})$ is complex
Gaussian distribution with mean $\textbf{a}$ and covariance matrix
$\textbf{A}$. $\textbf{E} [\cdot]$ stands for
expectation. $\mathbb{C}^{M\times N}$ is the space of complex $M\times
N$ matrices. $\lfloor \cdot \rfloor$ and $\lceil \cdot \rceil$
represent floor and ceiling operation, respectively.
$(\mathbf{a})_{i}$ indicates the $i$th element of the vector
$\mathbf{a}$ and $(\mathbf{A})_{ij}$ indicates the $ij$th element of
matrix $\mathbf{A}$.  
\section{Signal Model}\label{SecSigModl}
%

Consider a $K$-user $N_t \times N_r$ narrowband interference network
where there are $K$ user pairs and each transmitter and receiver are
equipped with $N_t$ and $N_r$ antennas respectively. Each transmitter
uses one DoF or sends one data stream at a time but there are $K$
simultaneous links from the $K$ transmitters within the same
band. Assume the channel is block-fading, i.e., the channel remains
the same within one frame and changes from one frame to
another. $\mathbf{H}_{k\ell}\in {\mathbb{C}}^{N_r\times N_t}$ is the
channel coefficient matrix between the transmitter $\ell$ and the
receiver $k$ of the considered frame (for clarity, we do not introduce
the frame index here). Each entity of $\mathbf{H}_{k\ell}$ is
independent and identically distributed (i.i.d.)
$\mathcal{{CN}}(0,1)$. The signal arriving at receiver $k$ is
\begin{align}
\mathbf{y}_{k} = \sum_{\ell=1}^{K}\mathbf{H}_{k\ell}{\textbf{v}}_{\ell} s_{\ell} +\mathbf{w}_{k}, \hspace{0.2in}  \text{for}\ k= 1,2\cdots ,K \label{eq:mdl}
\end{align}
where ${\textbf{v}}_{\ell}\in \mathbb{C}^{N_t \times 1}$ is the
unit-norm precoding vector associated with $s_{\ell}$, the transmitted signal of transmitter $\ell$. Each transmitter has a power constraint $P$, i.e., $\mathbf{E}[
  \left\| s_{\ell} \right \|^{2} ] = P$.
$\mathbf{w}_{k}\in\mathbb{C}^{N_r\times 1}$ is additive white Gaussian noise (AWGN) with
distribution
$\mathcal{{CN}}(\boldsymbol{0}, \boldsymbol{I})$.

At the receiver $k$, a unit-norm receive combining vector
$\textbf{u}_{k}\in\mathbb{C}^{N_r\times 1}$ is applied to suppress the
interference from other streams:
\begin{align} \label{eq:rxsignal}
{\textbf{u}}_{k}^{\dag} {\textbf{y}}_{k} & = \mathbf{{u}}_k^{\dag} \mathbf{H}_{kk} \textbf{{v}}_{k} s_{k}
 + \mathbf{{u}}_k^{\dag} \sum_{\ell=1,\ell\neq k}^{K} \mathbf{H}_{k\ell} {\textbf{v}}_{\ell}s_{\ell}  +\mathbf{{u}}_{k}^{\dag}\mathbf{w}_{k}.
\end{align}
Assuming interference alignment is feasible \cite{YJafar2010}, the alignment is achieved when the precoding and receive combining vectors satisfy:
\begin{align}
\mathbf{u}_{k}^{\dag}\mathbf{{H}}_{k\ell}\mathbf{v}_{\ell} & = 0,\forall \ell\neq k \label{eq:fw1}\\
\mathbf{u}_{k}^{\dag}\mathbf{{H}}_{kk}\mathbf{v}_{k}& \neq 0 \label{eq:fw3}, k = 1, \ldots, K.
\end{align}
Several algorithms \cite{GJafar2008,YSung2010,SPRH2009AlterMin} have
been proposed to solve for the precoding and receive combining
vectors. Among them, two of the most representative ones are the
iterative algorithms proposed by Gomadam, Cadambe and Jafar in
\cite{GJafar2008}: one aims to achieve perfect interference alignment
by minimizing the interference leakage (MinIL), the other one intends
to maximize the SINR of each user link (Max-SINR). In the following,
we use the two algorithms\footnote{We assume global CSI is available at transmitters.
The transmitters can perform the two IA algorithms and inform the receivers
their corresponding receiving combining vectors via control
information; or if global CSI is also available at the receivers, both transmitters and
receivers can perform the algorithms themselves and hence no extra control information is needed.
The CSI can be obtained by pilot and feedback signals, however, the specific mechanism of CSI exchange is
 out of the scope of this paper.} as examples to analyze the BER performance of IA.

\section{Transmission Schemes for Interference Channel and Their BER Analyses}
\label{SecBER}

\subsection{Minimum Interference Leakage Algorithm}\label{sub1}

At the receiver $k$, the total interference power caused by other
transmitters is
 \begin{align}\label{Lk}
 L_{k} = P \left( \mathbf{{u}}_k^{\dag} \sum_{\ell = 1, \ell \neq k}^{K} \mathbf{H}_{k\ell} \mathbf{{v}}_{\ell} \mathbf{{v}}_{\ell}^{\dag} \mathbf{H}_{k\ell}^{\dag} \mathbf{{u}}_k \right ).
 \end{align}
In MinIL algorithm, the precoding vectors $\{\mathbf{v}_k\}$ and the
receive combining vectors $\{\mathbf{u}_k\}$ are designed to force
$\{L_k\}$ to be zero~\cite{GJafar2008}, i.e., satisfying
condition~\eqref{eq:fw1}, so that the interference at each receiver is
completely eliminated.
Therefore, the post-processing received signal (\ref{eq:rxsignal}) can
be rewritten as
\begin{align}
{\textbf{u}}_{k}^{\dag} {\textbf{y}}_{k} & = \mathbf{{u}}_k^{\dag} \mathbf{H}_{kk} \textbf{{v}}_{k} s_{k} +(\mathbf{{u}}_k)^{\dag}\mathbf{w}_{k}.
 \label{eq:rxsignal2}
\end{align}
Denote ${z}_{k} \triangleq \mathbf{{u}}_k^{\dag} \mathbf{H}_{kk}
{\textbf{v}}_{k} $, the effective channel between user pair $k$, is
non-zero with probability~1~\cite{GJafar2008}. Therefore, in MinIL
algorithm, the design of $\{\mathbf{u}_k \}$ and
$\{\mathbf{v}_{\ell}\}$ actually focuses on condition~\eqref{eq:fw1}
and does not involve the direct channel $\{\mathbf{{H}}_{kk}\}$. Thus,
$\mathbf{u}_k$ and $\mathbf{v}_k$ are independent of
$\mathbf{{H}}_{kk}$. Conditioned on $\{\mathbf{H}_{k\ell}\}_{\ k \neq
  \ell}$, and hence $\mathbf{u}_k$ and $\mathbf{v}_k$, ${z}_{k}$ is a
complex Gaussian random variable with zero mean and variance:
 \begin{align}
 \label{fz}
 \textbf{E} \left [  \left | z_{k} \right |^2  \right ]  & = \textbf{E} \left [  \sum_{i = 1}^{M} \sum_{j = 1}^{M} |(\mathbf{{u}}_k)_{i} |^{2} |(\mathbf{{v}}_k)_{j} |^{2} |(\mathbf{H}_{kk})_{ij}|^{2}  \right ]\\\nonumber
  &= \sum_{i = 1}^{M} |(\mathbf{{u}}_k)_{i} |^{2} \sum_{j = 1}^{M} |(\mathbf{{v}}_k)_{j} |^{2}  \textbf{E} \left [  |(\mathbf{H}_{kk})_{ij}|^{2}  \right ] = 1,
  \end{align}
 where the last step holds because $\mathbf{{H}}_{kk}$ are
 i.i.d. $\mathcal{{CN}}(0, 1)$, and $\mathbf{u}_k$ and
 $\mathbf{v}_{k}$ have unit norm. So $\left |{z}_{k} \right |^2$ is an
 exponentially distributed random variable with unit mean and the
 effective channel under MinIL is Rayleigh. In other words, MinIL
 effectively decomposes the $K$-user interference network into $K$
 equivalent SISO Rayleigh fading channels.

At the receiver $k$, the post-processing SINR is
\begin{align} \label{SINRIA}
\gamma_{k} = {\left |z_{k}\right |^2 P} .
\end{align}
From (\ref{SINRIA}) and the exact BER expression of $M=I \times J$
rectangular QAM in~\cite{ExactBER2002}, the BER of IA with MinIL
algorithm is
\begin{align}
\label{BER_AWGN}
\mathrm{BER}(\left | z_{k} \right |)= \frac{1}{\log_2(I\cdot J)}\left ( \sum_{m = 1}^{\log_2 I}P_{I}(m) + \sum_{n = 1}^{\log_2 J}P_{J}(n) \right )
\end{align}
where
\begin{align}\label{PI}
P_{I}(m)  = \frac{2}{I}\sum_{i = 0}^{(1-2^{-m})I -1} \left \{ \eta(i,m,I)(-1)^{\left \lfloor \frac{i\cdot 2^{m-1}}{I} \right \rfloor } \right . \nonumber\\
 \left. \times Q \left ( \sqrt{\frac{6 (2i+1)^2 \left |z_{k}\right |^2 P}{\left (I^2 + J^2 -2 \right ) }} \right ) \right \}
\end{align}
\begin{align}\label{PJ}
P_{J}(n)  = \frac{2}{J}\sum_{i = 0}^{(1-2^{-n})J -1} \left \{ \eta(i,n,J)(-1)^{\left \lfloor \frac{i\cdot 2^{n-1}}{J} \right \rfloor } \right . \nonumber\\
 \left. \times Q \left (  \sqrt{\frac{6 (2i+1)^2 \left |z_{k}\right |^2 P}{\left (I^2 + J^2 -2 \right ) }}\right )\right \}
\end{align}
with $\eta(i,m,I) = \left ( 2^{m-1} - \left \lfloor \frac{i\cdot 2^{m-1}}{I}  + \frac{1}{2}\right \rfloor \right )$.
Since $\left | z_{k} \right |$ has Rayleigh distribution, we have
\begin{align}\label{BERIA}
\overline{{\mathrm{BER}}}_{\mathrm{IA}} & = \int_{0}^{+\infty} \mathrm{BER}(\left | z_{k} \right |) \times 2 \left |z_{k}\right | e^{-\left |z_{k} \right |^{2}} d\left | z_{k} \right |\\\nonumber
 & = \frac{1}{\log_2(I\cdot J)}\left ( \sum_{m = 1}^{\log_2 I}\overline{P}_{I}(m) + \sum_{n = 1}^{\log_2 J}\overline{P}_{J}(n) \right )
\end{align}
where
\begin{align}\label{PI2}
\overline{P}_{I}(m)  = &\frac{1}{I}\sum_{i = 0}^{(1-2^{-m})I -1} \left \{ \eta(i,m,I)(-1)^{\left \lfloor \frac{i\cdot 2^{m-1}}{I} \right \rfloor } \right . \nonumber\\
& \left. \times \left ( 1- \left ( \frac{\left (I^2 + J^2 -2 \right )  }{3 (2i+1)^2 P} + 1 \right )^{-1/2} \right )
\right \}
\end{align}
\begin{align}\label{PJ2}
\overline{P}_{J}(n)  = &\frac{1}{J}\sum_{i = 0}^{(1-2^{-n})J -1} \left \{ \eta(i,n,J) (-1)^{\left \lfloor \frac{i\cdot 2^{n-1}}{J} \right \rfloor } \right . \nonumber\\
& \left. \times \left ( 1- \left ( \frac{\left (I^2 + J^2 -2 \right ) }{3 (2i+1)^2 P} + 1 \right )^{-1/2} \right )
\right \}.
\end{align}

As a specific example, consider 4-QAM (QPSK), which leads to
\begin{align}
\overline{{\mathrm{BER}}}_{\mathrm{IA}} = \frac{1}{2} \left ( 1 - \sqrt{\frac{P}{P + 2  }} \right )\approx \frac{1}{2P}.
\label{BERIAQPSKAPP}
\end{align}
This shows that the diversity order of IA with MinIL is one,
which is consistent with the result in
\cite{IADiverisity2011}. Intuitively, the MinIL algorithm does not have
either diversity gain or array gain, since the precoding and receive combining
vectors are independent of the direct channel,
which leads to the equivalent channel uniformly distributed in an
interference-free subspace.


\subsection{Max-SINR Algorithm}\label{sub2}

The MinIL algorithm is suboptimal for low and intermediate SNR, because it
only focuses on eliminating interference and does not consider the
desired signal power. To improve the performance for low and
intermediate SNR, the Max-SINR algorithm is proposed~\cite{GJafar2008},
where each transceiver pair maximizes its corresponding SINR instead
of merely suppressing interference. 
In the Max-SINR algorithm, the
precoding vectors $\{\mathbf{v}_\ell\}$ and the receive combining vectors
$\{\mathbf{u}_k\}$ are designed in an iterative manner, so that the
instantaneous SINR of the $k$th pair
\begin{align}\label{SINR_MAXSINR}
\mathrm{SINR}_{k} = \frac{P \left |
  \mathbf{u}_{k}^{\dag}\mathbf{H}_{kk}\mathbf{v}_{k}\right |^2}{1 +
  L_k}, \ 1\le k\le K
\end{align}
is maximized~\cite{GJafar2008} where $L_k$ is defined in \eqref{Lk}.

An exact BER analysis of the Max-SINR algorithm is
intractable~\cite{IADiverisity2011} because $\{\mathbf{u}_k\}$ and
$\{\mathbf{v}_\ell\}$ depend on $\{ \mathbf{H}_{kk}\}$, and the
algorithm is iterative. Here, we provide an approximate analysis of the
SINR achieved by this algorithm. First, unlike the MinIL algorithm, in
the Max-SINR algorithm the interference $L_k$ is not necessarily zero,
but it is bounded as $P$ goes to infinity. Intuitively, this can
be verified as follows: If $L_k=f(P)$ where
$\lim_{P\rightarrow \infty}f(P)=\infty$, then the SINR is
{\em not} maximized, since simply forcing $L_k$ to zero leads to
higher SINR for sufficiently large $P$. Thus, we approximate the
residual interference as a complex Gaussian random variable with
bounded variance $\delta$. Then, the post-processing SINR can be
rewritten as
\begin{align}\label{gamma_MAXSINR}
\gamma_k^{\mathrm{M}} = \frac{P \left | z_{k}^{\mathrm{M}} \right |^2}{1+\delta}
\end{align}
where $z_{k}^{\mathrm{M}}=
\mathbf{{u}}_{k}^{\dag}\mathbf{H}_{kk}\mathbf{{v}}_{k}$ is the
equivalent channel of the $k$th user pair under the Max-SINR
algorithm. On the one hand, the Max-SINR algorithm searches for
precoding/receive combining vectors that lead to larger desired signal
power relative to the MinIL algorithm, and hence a coherent combining
gain is expected. On the other hand, the coherent combining gain of
$|z_{k}^{\mathrm{M}}|$ is upper bounded by that achieved by
beamforming~\cite{STBCbook2003}, since the Max-SINR also needs to
suppress the interference.

\begin{figure}
\centering
\includegraphics[width=3.7in]{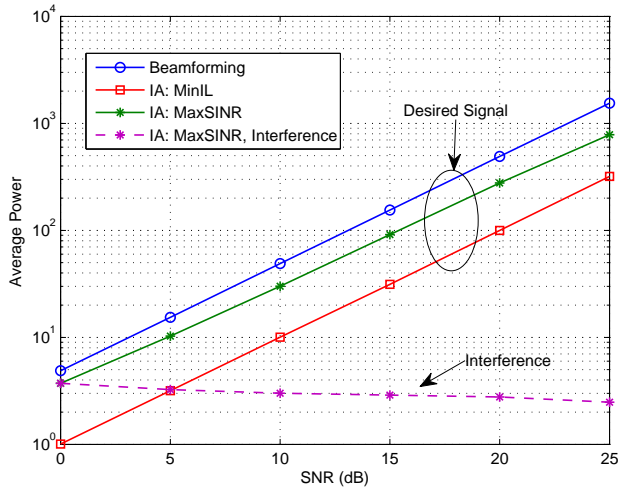}
\caption{Average power of desired signal and interference: $K = 3, N_t = 3, N_r =2$}
\label{MEAN_PD_PI_NT3}
\end{figure}

To support the above analysis, we provide numerical results for the
average desired signal power ($| z_{k}^{\mathrm{M}}|^2$), and
interference power ($L_k$) in Fig.~\ref{MEAN_PD_PI_NT3}, where a
3-user ($3\times 2$) interference network is considered. One can see
that the residual interference power is bounded (not growing with
$P$), and in fact the interference is at a similar power level as
noise. In addition, the average power of the desired signal lies in
between the powers of those in MinIL and beamforming.


\subsection{Spatial Multiplexing}
\label{sub3}

We consider spatial multiplexing (SM) as a benchmark. Here, $K$
transceiver pairs are scheduled in a time-division manner (TDMA). When
the $k$th pair is activated, its corresponding channel is decomposed as
\begin{align}
\mathbf{{H}}_{kk} = \mathbf{USV}^{\dag}
\end{align}
where $\mathbf{V}$ and $\mathbf{U}$ are unitary precoding matrix and
receive combining matrix, and
\begin{align}
 \mathbf{S} = \mathrm{diag} [\lambda_1, \cdots, \lambda_{N_{\mathrm{min}}},\underbrace{0,\cdots, 0}_{N_{\mathrm{max}}-N_{\mathrm{min}}}].
\end{align}
In the above, $N_{\mathrm{min}} = \min(N_t,N_r)$ and $N_{\mathrm{max}}
= \max(N_t,N_r)$, and $\lambda_1 \ge \cdots \ge
\lambda_{N_{\mathrm{min}}}$ are singular values. In this case, the
MIMO channel of this user pair is decomposed into $N_{\mathrm{min}}$ parallel SISO links
(multiplexing gain $N_{\mathrm{min}}$).

To communicate at the same rate and power as IA, spatial
multiplexing must have constellation size $2^{K
  R_{\mathrm{IA}}/N_{\mathrm{min}}}$ and power $KP$, where
$R_{\mathrm{IA}}$ is the rate used by IA per transceiver pair. The
exact BER expression of SVD-based MIMO system is given by
\cite{BERSVD2006}; we rewrite it in a general form that is applicable
to $I \times J$ rectangular QAM:
\begin{align}
\overline{{\mathrm{BER}}}_{\mathrm{SVD}} = \frac{1}{N_{\mathrm{min}} \log_2(I\cdot J)}\left ( \sum_{m = 1}^{\log_2 I}\widehat{P}_{I}(m) + \sum_{n = 1}^{\log_2 J}\widehat{P}_{J}(n) \right )
\end{align}
where
\begin{align}
 &\widehat{P}_{I}(m)= \frac{1}{I}\sum_{p = 1}^{\log_2 I}\sum_{c = 0}^{(1-2^{-p})I -1} \sum_{i = 1}^{N_{\mathrm{min}}} \sum_{m= 0}^{i-1}\sum_{j = m+\Delta}^{m+\Delta+i-1} \sum_{b=0}^{j}  \nonumber\\
& \frac{\left (-1 \right )^{2j-\Delta -b + \left \lfloor \frac{c\cdot 2^{p-1}}{I}  + \frac{1}{2}\right \rfloor }(i-1)!b! }{\left( i-1+\Delta \right)! \left ( j-\Delta \right )!}
{j-\Delta \choose m}{i-1+\Delta \choose i-1-m}\nonumber\\
&{i-1+\Delta \choose j-m} {j \choose b}
\eta(c,p,I)  \left ( 1- \sum_{k=0}^{b} {2k \choose k} \frac{\beta_{c}^{\frac{1}{2}}}{2^k (\beta_c + 2)^{k+\frac{1}{2}}} \right )\label{PI2SVD}
\end{align}
with $\Delta = N_{\mathrm{max}} - N_{\mathrm{min}}$ and $\beta_c = \frac{6(2c+1)^2 KP}{\left (I^2 + J^2 -2 \right )
  N_{\mathrm{min}}}$. By substituting $I$ with $J$ in (\ref{PI2SVD}),
$\widehat{P}_{J}(m)$ can be obtained.

To gain some insights, we consider $N_t = N_r = N$. Note that the BER
performance of spatial multiplexing is essentially dominated by the
smallest eigenmode.  Since the minimum eigenvalue of a Wishart matrix
here is exponentially distributed with parameter
$N$~\cite{STBCbook2003}\cite{Edelman1988}, the smallest singular value
${\lambda_{\mathrm{min}}}$ is Rayleigh distributed with pdf
$f_{\lambda_{\mathrm{min}}}(\lambda) = N\lambda e^{-N\lambda^2}$. For
4-QAM, at high SNR the average BER can be approximated as
\begin{align}  \label{BERSVDQPSKAPP}
\overline{{\mathrm{BER}}}_{\mathrm{SVD}} &\approx  \int_{0}^{+\infty} Q \left (\sqrt{\frac{\lambda^2PK}{N }} \right ) f_{\lambda_{\mathrm{min}}}(\lambda) d\lambda \\\nonumber
 & = \frac{1}{2} \left ( 1 - \sqrt{\frac{PK}{PK+ 2N^2}} \right ) \approx \frac{N^2}{2PK}.
\end{align}
One can see that the diversity order is also one in this case.

\subsection{Performance Comparison}\label{sub4}
\begin{table*}
\caption{Performance comparison of a $K$-user $N_t \times N_r$ MIMO
  interference network}
\label{Table}
\centering
\begin{tabular}{|c|c|c|c|}
\hline
 & Multiplexing Gain & Diversity Order & Coherent Combining Gain\\
\hline
IA with MinIL & $KN_{\mathrm{min}}/2$ & 1 & 1\\
\hline
\multirow{2}{*}{IA with Max-SINR} & \multirow{2}{*}{ $KN_{\mathrm{min}}/2$ } & 1, for $ N_t + N_r = K +1$  & \multirow{2}{*}{$> 1, < \mathbf{E}[\left |\lambda_1 \right |^2]$} \\
\cline{3-3}
 &  & $\geq 1$, for $N_t + N_r \geq K +2$& \\
 \hline
SM with SVD & $N_{\mathrm{min}}$ &  $N_{\mathrm{max}} - N_{\mathrm{min}} +1$ & Depends on singular value distribution \\
\hline
\end{tabular}
\end{table*}

In this subsection, we summarize and compare the multiplexing gains,
diversity orders and the coherent combining gains of the
above-mentioned three transmission strategies.

From a capacity perspective, the multiplexing gain of IA is
$\frac{KN_{\mathrm{min}}}{2}$ for both MinIL and Max-SINR algorithms
whereas SM achieves $N_{\mathrm{min}}$ multiplexing gain. Clearly,
when there are more than two user pairs in an interference network, IA
has a larger asymptotic capacity.

MinIL has performance similar to a SISO channel under Rayleigh fading
and offers neither diversity nor array gain.
Max-SINR maximizes the SINR for every user pair and obtains a coherent
combining gain.  It also achieves a diversity gain when $N_t + N_r
\geq K+2$~\cite{IADiverisity2011} since the extra dimensions enable
precoding and receive combining vectors to be selected from a group of
candidates.

For SVD-based SM, the diversity order is one when $N_t=N_r$, and is
$N_{\mathrm{max}} - N_{\mathrm{min}} + 1$ when $N_t\neq
N_r$~\cite{SVDDiversity2006}. The expression for the array gain is
complicated and determined due to its dependence on the distributions
of the singular values~\cite{BERSVD2006}.

The above results are summarized in Table~\ref{Table}.  MinIL and
Max-SINR share the same multiplexing gain but the latter also enjoys
coherent combining gain and possible diversity gain. Therefore,
Max-SINR outperforms MinIL. However, since the diversity order and the
coherent combining gain of SVD-based SM depend on the MIMO channel
configuration~\cite{BERSVD2006,SVDDiversity2006}, the comparison between IA and SM does
not produce a definitive conclusion. Further comparisons will be
presented in Section \ref{SecSimulation} via simulations.

\section{Bit Loading and adaptive transmission scheme}\label{SecBL}
\subsection{Bit Loading}\label{subBL}

Both interference alignment and SVD-based spatial multiplexing provide
a number of equivalent channels or ``pipes'' for communication. When
CSIT is available, one can exploit it by applying adaptive bit loading
to those equivalent channels to reduce the error rate or to increase
the data rate. In our approach, for data rate $R$, equal power $KP/R$
is allocated to each bit.

For  MinIL, the bit loading is:
\begin{align}\label{optIA}
\left (M_{1}^{\star}, \cdots, M_{{K}}^{\star} \right ) &= \arg \min_{\left (M_{1}, \cdots, M_{K} \right )} \frac{1}{R} \sum_{i = 1}^{K} P_b(i,M_i,{z}_i)\log_2 (M_i)\\\nonumber
s.t.\hspace{0.2in} R &= \sum_{n = 1}^{K}\log_2 (M_n),\ \
 0 \leq  M_n \leq 2^R.
\end{align}
$P_b(i,M_i,z_{i})$, the instantaneous bit error probability of the
$i$th equivalent channel, is obtained by substituting the modulation
size $M_i$ and equivalent SINR $|z_{i}|^2 KP/R \log_2(M_i)$ for
$|z_{k}|^2 P$ in the general $M_i$-QAM BER expression for AWGN
channels (\ref{BER_AWGN}-\ref{PJ}). The above optimization problem can
be easily solved by the iterative algorithm described in
Table~\ref{BLTable}, which terminates in $R$ steps.

\begin{table}
\caption{Iterative bit loading algorithm}
 ----------------------------------------------------------------------------------------------------------------------------------------------------------
\begin{enumerate}
  \item Initialize the bit allocation scheme as $\left [ M_{1}, \cdots, M_{K} \right ] := \left [1, \cdots, 1 \right ]$.
  \item Add one bit and its associated power to $i$th  equivalent channel while the modulation sizes of other channels remain the same:
  \begin{align}
  & \left [ \log_2 (M_{1}), \cdots, \log_2 (M_{i}) \cdots, \log_2 (M_K) \right ] \\
  &= \left [ \log_2 (M_{1}), \cdots, \log_2 (M_{i}) +1 \cdots, \log_2 (M_K) \right ], \nonumber
  \end{align}
  for $\ i = 1, \cdots, K$.
  Totally, there are $K$ ways to add the additional bit which are enumerated row-wisely in the following matrix:
    \[
    \begin{pmatrix}
     \log_2 (M_1) + 1  & \log_2 (M_{2}) & \cdots& \log_2 (M_{K})\\
     \log_2 (M_1)  & \log_2 (M_{2}) + 1 & \cdots& \log_2 (M_{K})\\
     \vdots & \vdots & \ddots &\vdots \\
     \log_2 (M_1)  & \log_2 (M_2)&\cdots& \log_2 (M_{K})+ 1
    \end{pmatrix}
    \]
  \item Compute the effective BERs of the above $K$ strategies by a weighted sum $ \frac{1}{R}{\sum_{i = 1}^{K} P_b(i,M_i,z_i)\log_2 (M_{i})}$. Choose and update the allocation strategy with the one which returns the minimum weighted BER.
  \item Check if $ \sum_{i = 1}^{K} \log_2(M_i) = R$, if not, go back to 2), otherwise, return the current $\left [ M_{1}, \cdots, M_K \right ]$ .
\end{enumerate}
 \label{BLTable}
 ----------------------------------------------------------------------------------------------------------------------------------------------------------
\end{table}

After bit loading, the power transmitted over each equivalent channel
may change. However, the directions of $\{\mathbf{u}_k\}$ and
$\{\mathbf{v}_\ell\}$ still satisfy conditions (\ref{eq:fw1}) and
(\ref{eq:fw3}). Therefore, $\{\mathbf{u}_k\}$ and
$\{\mathbf{v}_\ell\}$ are unaffected.

The bit loading procedure for Max-SINR follows in a similar manner as
MinIL:
\begin{align}\label{optIA_MAXSINR}
&\hspace{0.9in} \left (M_{1}^{\star}, \cdots, M_{{K}}^{\star} \right )\nonumber\\ &= \arg \min_{\left (M_{1}, \cdots, M_{K} \right )} \frac{1}{R}\sum_{i = 1}^{K} P_b^{'}(i,M_i,{\mathrm{SINR}}_i)\log_2 (M_i)\\\nonumber
& s.t.\hspace{0.2in} R= \sum_{n = 1}^{K}\log_2 (M_n),\ \
 0 \leq  M_n \leq 2^R,
\end{align}
where $P_b^{'}(i,M_i,{\mathrm{SINR}}_i)$ is the BER of the equivalent
channel $i$, obtained similarly as the corresponding value
in~\eqref{optIA} except using $\mathrm{SINR}_i$
from~\eqref{SINR_MAXSINR}. To do so a Gaussian assumption is used on
interference, which is an approximation.

Unlike MinIL, in Max-SINR the precoding and receive combining vectors
must be re-designed after bit loading. This is because the Max-SINR
transmit and receive vectors depend on the power allocated to
equivalent channels. Our approach is to apply Max-SINR (with updated
bit/power allocation) once again\footnote{Further iterations do not
  produce appreciable gain.} to obtain a new set of $\{\mathbf{u}_k\}$ and
$\{\mathbf{v}_\ell\}$.

The bit loading for SVD-based SM is performed similarly by replacing
$K$ with ${N_{\mathrm{min}}}$:
\begin{align}\label{optSM}
&\hspace{0.9in} \left (M_{1}^{\star}, \cdots, M_{N_{\mathrm{min}}}^{\star} \right ) \nonumber\\
&= \arg \min_{\left (M_{1}, \cdots, M_{N_{\mathrm{min}}} \right )} \frac{1}{R}\sum_{i = 1}^{N_{\mathrm{min}}} P_b(i,M_i,\lambda_i)\log_2 (M_{i})\\\nonumber
& s.t.\hspace{0.2in} R= \sum_{n = 1}^{N_{\mathrm{min}}} \log_2 (M_n),\ \
 0 \leq M_n \leq 2^R.
\end{align}
The diversity order of SVD-based SM after the above bit loading is
$N_tN_r$, since the bit loading algorithm for SM includes beamforming
as a special case~\cite{STBCbook2003}. For example, in the low-SNR
regime it allows the transmitter to spend all the power in the dominant
eigenmode.
\subsection{Adaptive Transmission Scheme}\label{subsecAdap}
As will be shown in Section~\ref{SecSimuBL}, the bit loading algorithm
significantly reduces the error rates of MinIL, Max-SINR and SVD-based
SM.  Since the performance of none of these schemes dominates at all
SNR, one may consider an adaptive scheme so that for each channel
condition, the best of the three is selected and used. This requires
knowledge of CSI, but the three schemes already assume existence of
CSI, so no further assumptions are introduced. To summarize, an
adaptive transmission scheme will be designed which can switch among
MinIL, Max-SINR and SVD-based SM, all with bit loading.

We first optimize each of the three transmission modes according to
the CSI. Then all users select a single transmission mode which
achieves the minimum BER. The adaptation rate is the same as the rate
of CSI update.  To elaborate, for each transmission mode, bit loading is
applied to obtain the optimal constellation vectors $\{M_i^\star\}$ as
described in the previous subsection. This also provides the
corresponding average BER of all users at the given CSI for each
mode. We then select the mode with the lowest BER:
\begin{align}
m^{\star}= &\arg \min_{m \in \{\mathrm{IA: MinIL,\ IA: Max-SINR,\ SM: SVD}\} } \nonumber \\
&\{P_{\mathrm{BL}}^{\mathrm{MinIL}}, P_{\mathrm{BL}}^{\mathrm{SVD}}, P_{\mathrm{BL}}^{\mathrm{Max-SINR}}\}
\end{align}
where
\begin{align}\label{PBLSVD}
P_{\mathrm{BL}}^{\mathrm{SVD}} \triangleq \frac{1}{R}{\sum_{i = 1}^{N_{\mathrm{min}}} P_b(i,M_{i}^{\star},{\lambda}_i) \log_2 (M_{i}^{\star})},
\end{align}
\begin{align}\label{PBLMIN}
P_{\mathrm{BL}}^{\mathrm{MinIL}} \triangleq \frac{1}{R}{\sum_{i = 1}^{K} P_b(i,M_{i}^{\star},{z}_i)\log_2 (M_{i}^{\star})},
\end{align}
 \begin{align}\label{PBEL_MAXSINR}
 P_{\mathrm{BL}}^{\mathrm{Max-SINR}}\triangleq \frac{1}{R}{\sum_{i = 1}^{K} P_{b}^{'}(i,M_{i}^{\star},{\mathrm{SINR}}_{i}^{\star})\log_2 (M_{i}^{\star})}.
 \end{align}
Note that the selected mode $m^{\star}$ is the same for all users and
remains the same during one frame. The corresponding bit allocations
$\{M_i^\star\}$ in \eqref{PBLSVD}-\eqref{PBEL_MAXSINR} could be
different across $K$ users.

The details of calculation of the bit-loading information as well as
the receive and transmit filters can vary according to the system
requirements. If, for example, the transmitters are base stations and
the receivers are mobiles (downlink), it is reasonable that the global
CSI is aggregated at the transmitter where calculations are also made,
and communicated with the receivers. In an uplink scenario, the
situation would be reversed.

The performance of the adaptive transmission scheme will be presented
in Section \ref{SecSimulation}.

\section{CSI Uncertainty}
\label{SecCSI}
In practice, perfect CSI may not be available, therefore one is often
interested in the performance of the communication system under
partial CSI. We model the channel
as~\cite{EtkinTse2006DOF}:
\begin{align}\label{CHMODEL}
\mathbf{H}_{k\ell} =\sqrt{1- \epsilon} \mathbf{\hat{H}}_{k\ell} + \sqrt{\epsilon}\mathbf{W}_{k\ell}
\end{align}
where $\mathbf{H}_{k\ell}$ is the channel between the transmitter
$\ell$ and the receiver $k$, which is known by the receiver $k$,
$\mathbf{\hat{H}}_{k\ell} \in {\mathbb{C}}^{N_r\times N_t}$ is the
channel {\em estimate} known by all the transmitters, and
$\mathbf{W}_{k\ell}$ is related to the channel error matrix and is
independent of $\mathbf{\hat{H}}_{k\ell}$. All elements of
$\mathbf{\hat{H}}_{k\ell}$ and $\mathbf{W}_{k\ell}$ are
i.i.d. $\mathcal{{CN}}(0,1)$. The rationale behind assuming perfect
channel state information at receiver (CSIR) but imperfect CSIT is
because it is always easier to obtain CSIR relative to CSIT.\footnote{
  For example, the receivers first estimate the CSI and feed it back
  to transmitters. Due to limited feedback or feedback delay, the
  transmitters only have the access to partial CSI
  $\mathbf{\hat{H}}_{k\ell}$ whereas the receivers always know the
  actual channel $\mathbf{{H}}_{k\ell}$ (since they can keep on
  estimating the CSI).} The parameter $0\le \epsilon \le 1$ reflects
uncertainty of CSIT: $\epsilon = 0$ corresponds to perfect CSIT
whereas $\epsilon = 1$ means that CSIT is completely unreliable.

In the following, we will first discuss the effects of CSI uncertainty
based on equal power allocation, and then describe the bit loading and
adaptive transmission scheme under CSI uncertainty in the last
subsection.

\subsection{Minimum Interference Leakage Algorithm}
Based on available CSIT, the transmitters design the precoding and
receive combining vectors to satisfy
\begin{align}
\mathbf{\hat{u}}_{k}^{\dag}\mathbf{\hat{H}}_{k\ell}\mathbf{\hat{v}}_{\ell} & = 0,\forall \ell\neq k \label{eq:fw2}\\
\mathbf{\hat{u}}_{k}^{\dag}\mathbf{\hat{H}}_{kk}\mathbf{\hat{v}}_{k}& \neq 0 \label{eq:fw4}, k = 1, \ldots, K.
\end{align}
In this case, there is residual interference due to imperfect CSI and the resulting
received signal is:
\begin{align} \label{}
{\mathbf{\hat{u}}}_{k}^{\dag} {\mathbf{y}}_{k} & = \mathbf{\hat{u}}_k^{\dag} \mathbf{H}_{kk} \mathbf{\hat{v}}_{k} s_{k}
 + \sqrt{\epsilon} \mathbf{\hat{u}}_k^{\dag} \sum_{\ell=1,\ell\neq k}^{K} \mathbf{W}_{k\ell} {\mathbf{\hat{v}}}_{\ell}s_{\ell}  +(\mathbf{\hat{u}}_k)^{\dag}\mathbf{w}_{k}.
\end{align}
Denote the equivalent channel as $\hat{z}_{k} =
\mathbf{\hat{u}}_{k}^{\dag}\mathbf{H}_{kk}\mathbf{\hat{v}}_{k}$. The
instantaneous SINR of the $k$th user pair is
\begin{align}\label{r1}
{\mathrm{SINR}}_{k} &= \frac{P \left | \hat{z}_{k} \right |^2 }{ \left ( 1 + P \sum_{\ell=1, \ell \neq k}^{K} \left |\mathbf{\hat{u}}_{k}^{\dag} \mathbf{H}_{k\ell}\mathbf{\hat{v}}_{\ell} \right |^2 \right )} \nonumber\\
&= \frac{P \left |\hat{z}_{k} \right |^2 }{ \left ( 1 +  \epsilon P \sum_{\ell=1, \ell \neq k}^{K} \left |\mathbf{\hat{u}}_{k}^{\dag} \mathbf{W}_{k\ell}\mathbf{\hat{v}}_{\ell} \right |^2 \right )}
\end{align}
Since both $\mathbf{\hat{u}}_k$ and $\mathbf{\hat{v}}_k$ are
independent of $\mathbf{\hat{H}}_{kk}$ (as mentioned in
Section~\ref{sub1}) and $\mathbf{{H}}_{kk}$ (based on the channel
model \eqref{CHMODEL}), $\hat{z}_{k}$ has the same distribution as
${z}_{k}$, and the corresponding analysis can be adopted for $\hat{z}_{k}$.
In addition,
$\mathbf{\hat{u}}_k$ and $\mathbf{\hat{v}}_k$ are independent of
$\mathbf{{W}}_{kk}$, and hence the interference power of the
data stream $k$ is
\begin{align}\label{Ik}
\hat{L}_{k} &= \epsilon P \sum_{\ell=1,\ell\neq k}^{K} \mathbf{E} \left [\left | \mathbf{\hat{u}}_k^{\dag} \mathbf{W}_{k\ell} {\mathbf{\hat{v}}}_{\ell} \right |^2 \right ]\nonumber\\
 &= \epsilon P \sum_{\ell=1,\ell\neq k}^{K} \mathbf{{E}} \left [ \sum_{i}^{M}|(\mathbf{\hat{u}}_k)_{i} |^{2} \sum_{j}^{M} | (\mathbf{\hat{v}}_{\ell})_{j} |^{2} |(\mathbf{W}_{k\ell})_{ij}|^{2} \right ]\nonumber\\
 & = \epsilon (K-1) P,
\end{align}
which suggests that the interference power grows with the transmit
power $P$, the number of users in the network $K$, and the level of
CSI uncertainty $\epsilon$. As a result, the post-processing SINR can be
written as
\begin{align}\label{r2}
\hat{\gamma}_{k} = \frac{\left |\hat{z}_{k}\right |^2 P} {\epsilon (K-1) P   + 1}.
\end{align}
Note that~\eqref{r2} reduces to (\ref{SINRIA}) if $\epsilon = 0$.

When CSIT is imperfect, with the assumption of Gaussian interference,
the BER expression of MinIL can be represented by (\ref{BERIA}) but
with modified $\overline{P}_{I}(m) $ and $\overline{P}_{J}(n)$:
\begin{align}
&\overline{P}_{I}(m)  = \frac{1}{I}\sum_{i = 0}^{(1-2^{-m})I -1} \left \{ \eta(i,m,I) (-1)^{\left \lfloor \frac{i\cdot 2^{m-1}}{I} \right \rfloor } \right.
\nonumber\\ & \left. \left ( 1- \left ( \frac{\left (I^2 + J^2 -2 \right ) \left ( \epsilon (K-1) P +1 \right ) }{3 (2i+1)^2 P} + 1 \right )^{-1/2} \right )
\right \}
\label{PI3}
\end{align}
\begin{align}
&\overline{P}_{J}(n)  =\frac{1}{J}\sum_{i = 0}^{(1-2^{-n})J -1} \left \{ \eta(i,n,J) (-1)^{\left \lfloor \frac{i\cdot 2^{n-1}}{J} \right \rfloor } \right.
\nonumber\\ & \left. \left ( 1- \left ( \frac{\left (I^2 + J^2 -2 \right ) \left ( \epsilon (K-1) P +1 \right ) }{3 (2i+1)^2 P} + 1 \right )^{-1/2} \right )
\right \}.
\label{PJ3}
\end{align}

\subsection{Max-SINR Algorithm}
For Max-SINR with imperfect CSIT, the precoding $\{ \mathbf{\hat{u}}_k
\}$ and receive combining vectors $\{ \mathbf{\hat{v}}_k\}$ are designed to
maximize
\begin{align}\label{rhat}
\hat{\mathrm{SINR}}_{k} = \frac{P \left | \mathbf{\hat{u}}_{k}^{\dag}\mathbf{\hat{H}}_{kk}\mathbf{\hat{v}}_{k} \right |^2 }{ \left ( 1 + P \sum_{\ell=1, \ell \neq k}^{K} \left |\mathbf{\hat{u}}_{k}^{\dag} \mathbf{\hat{H}}_{k\ell}\mathbf{\hat{v}}_{\ell} \right |^2 \right )}.
\end{align}
Using the above design, the actual equivalent channel is
\begin{align}
 \mathbf{\hat{u}}_{k}^{\dag}\mathbf{H}_{kk}\mathbf{\hat{v}}_{k} &= \sqrt{1-\epsilon}\mathbf{\hat{u}}_{k}^{\dag}\mathbf{\hat{H}}_{kk}\mathbf{\hat{v}}_{k} + \sqrt{\epsilon}\mathbf{\hat{u}}_{k}^{\dag}\mathbf{W}_{kk}\mathbf{\hat{v}}_{k} \nonumber\\
 &= \sqrt{1-\epsilon} \hat{z}_{k}^{\mathrm{M}} + \sqrt{\epsilon}\mathbf{\hat{u}}_{k}^{\dag}\mathbf{W}_{kk}\mathbf{\hat{v}}_{k}.
\end{align}
The actual interference power is
\begin{align}
\hat{L}_{k}^{\mathrm{M}} &=  P \sum_{\ell=1,\ell\neq k}^{K} \mathbf{E} \left [\left | \mathbf{\hat{u}}_k^{\dag} \mathbf{H}_{k\ell} {\mathbf{\hat{v}}}_{\ell} \right |^2 \right ]\\\nonumber
  &=  P \sum_{\ell=1,\ell\neq k}^{K} \mathbf{E} \left [\left |
  \sqrt{1-\epsilon}\mathbf{\hat{u}}_{k}^{\dag}\mathbf{\hat{H}}_{k\ell}\mathbf{\hat{v}}_{\ell}
  +
  \sqrt{\epsilon}\mathbf{\hat{u}}_{k}^{\dag}\mathbf{W}_{k\ell}\mathbf{\hat{v}}_{\ell}\right
  |^2 \right ] \nonumber\\
  & = (1-\epsilon) \hat{\delta} + \epsilon(K - 1)P.
\end{align}
The last step follows the analysis that led to (\ref{Ik}), since
$\mathbf{\hat{u}}_k$ and $\mathbf{\hat{v}}_{\ell}$ are independent of
$\mathbf{{W}}_{k\ell}$. Note that $\hat{z}_{k}^{\mathrm{M}} =
\mathbf{\hat{u}}_{k}^{\dag}\mathbf{\hat{H}}_{kk}\mathbf{\hat{v}}_{k}$
and $\hat{\delta}=P \sum_{\ell=1, \ell \neq k}^{K} \left
|\mathbf{\hat{u}}_{k}^{\dag}
\mathbf{\hat{H}}_{k\ell}\mathbf{\hat{v}}_{\ell} \right |^2$ have the
same distributions as ${z}_{k}^{\mathrm{M}}$ and $\delta$ in
(\ref{gamma_MAXSINR}), respectively. Therefore, the resulting actual
post-processing SINR is:
\begin{align}\label{rmm}
\hat{\gamma}_{k}^{\mathrm{M}} =  \frac{\left |
  \mathbf{\hat{u}}_{k}^{\dag}\mathbf{H}_{kk}\mathbf{\hat{v}}_{k}
  \right |^2 P} {\hat{L}_{k}^{\mathrm{M}} + 1} =
\frac{(1-\epsilon)\left |\hat{z}_{k}^{\mathrm{M}}\right |^2 P +
  \epsilon P} { (1-\epsilon)\hat{\delta} +\epsilon (K-1) P + 1}.
\end{align}

Compared with MinIL (see (\ref{r2})), in Max-SINR, CSI
uncertainty has a bigger impact on the post-processing SINR.  For
MinIL in (\ref{r2}), the CSI uncertainty causes a residual
interference of power $\epsilon( K- 1)P$ but does not reduce the
equivalent desired signal power. On the contrary, for Max-SINR in
(\ref{rmm}), CSI uncertainty affects two interference terms:
$\epsilon( K- 1)P$ and $(1-\epsilon)\hat{\delta}$. Moreover, the
average desired signal power is influenced by imperfect CSIT. The
desired signal power now splits into two parts----one comes from the
equivalent desired power $\left |\hat{z}_{k}^{\mathrm{M}}\right |^2$
based on CSIT but is scaled by $(1-\epsilon)$ and the other one
relates to the CSI imperfection. As such, Max-SINR  is
more sensitive to CSI uncertainty than MinIL.


\subsection{Spatial Multiplexing}
With CSI uncertainty, the exact BER expression of $I \times
J$ rectangular QAM in SVD-based MIMO system is given as below
\cite{BERSVD2006}:
\begin{align}
\overline{{\mathrm{BER}}}_{\mathrm{SVD}} = \frac{2 e^{\frac{\epsilon}{1-\epsilon}}}{N_{\mathrm{min}} \log_2(I\cdot J)}\left ( \sum_{m = 1}^{\log_2 I}\widehat{P}_{I}(m) + \sum_{n = 1}^{\log_2 J}\widehat{P}_{J}(n) \right )
\end{align}
where
\begin{align}\label{PI3SVD}
&\widehat{P}_{I}(m)= \frac{1}{I}\sum_{p = 1}^{\log_2 I}\sum_{c = 0}^{(1-2^{-p})I -1} \sum_{i = 1}^{N_{\mathrm{min}}} \sum_{m= 0}^{i-1}\sum_{j = m+\Delta}^{m+\Delta+i-1} \sum_{b=0}^{j} {j-\Delta \choose m} \nonumber \\
&\frac{\left (-1 \right )^{2j-\Delta -b + \left \lfloor \frac{c\cdot 2^{p-1}}{I}  + \frac{1}{2}\right \rfloor }(i-1)! \left (\frac{\epsilon}{1-\epsilon} \right )^{2(j-b)}}{\left( i-1+\Delta \right)! \left ( j-\Delta \right )!}{i-1+\Delta \choose i-1-m}\nonumber\\
&
{i-1+\Delta \choose j-m} {j \choose b}
\eta(c,p,I) \int_{\frac{\epsilon}{1-\epsilon} }^{\infty} Q\left (\sqrt{\beta_c \lambda} \right )\lambda^{b} e^{-\lambda}d\lambda
\end{align}
with a modified $\beta_c = \frac{6(2c+1)^2 (1-\epsilon)}{\left (I^2 + J^2 -2 \right ) \left (N_{\mathrm{min}}/(KP) + (N_{\mathrm{min}} -1 )\epsilon \right )}$. Similarly, $\widehat{P}_{J}(m) $ is obtained by replacing $I$ with $J$ in (\ref{PI3SVD}).
\subsection{Bit Loading and Adaptive Transmission Scheme}

In the presence of CSI uncertainty, the proposed bit loading
(Section~\ref{subBL}) and the adaptive algorithm
(Section~\ref{subsecAdap}) can still be applied, where all the
calculations and selection are based on the available CSIT. To fit in
the channel model in this section, $z_{i}$ is substituted by
$\hat{z}_{i}$ in (\ref{optIA}), $\mathrm{SINR}_{i}$ in
(\ref{optIA_MAXSINR}) is replaced by $\hat{\mathrm{SINR}}_{i}$ in
(\ref{rhat}) and $\lambda_{i}$ is substituted by $\hat{\lambda}_{i}$
in (\ref{optSM}) (where $\hat{\lambda}_{i}$ is the singular value of
$\mathbf{\hat{H}}_{kk}$). Although the CSIT is not perfect, the bit
loading and the adaptive transmission scheme still provide additional
gains compared with the non-bit-loaded case, as shown in
Section \ref{SecSimulation}.


\section{Simulation Results and Discussion}\label{SecSimulation}
In this section, the BER performances of IA with MinIL and Max-SINR as
well as the SVD-based SM are evaluated via Monte Carlo
simulations. This section is divided into two parts: the performances
with and without bit loading. For each part, we first focus on the
perfect CSIT cases and then discuss the effects of CSI uncertainty. In
our simulations, the transmit power of all the schemes are kept the
same. Moreover, the data rate is maintained the same among the three
transmission modes: IA transmits at rate $R_{\mathrm{IA}}$ per
transceiver pair, while SM transmits at rate $K R_{\mathrm{IA}}$. The
numbers of iterations of MinIL and Max-SINR are both set to
be 100.

\subsection{Performances of IA and SM without Bit Loading}
\subsubsection{Perfect CSIT}
\begin{figure}
\centering
\includegraphics[width=3.7in]{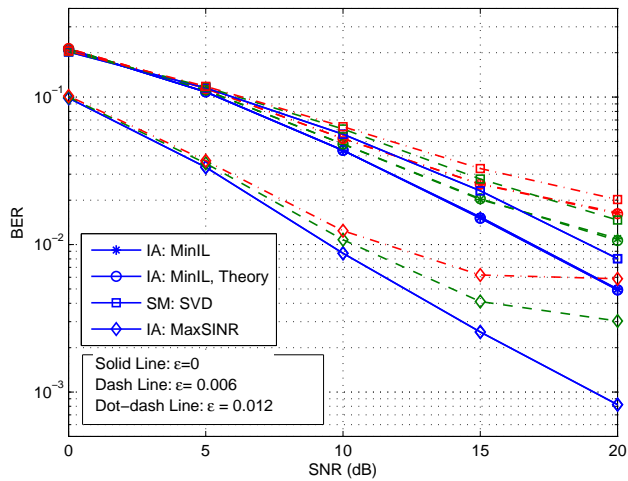}
\caption{BER performances of the three transmission modes without bit loading: $K = 3, N_t = N_r =2$; $\epsilon$ represents CSIT uncertainty: $\epsilon = 0$ corresponds to perfect CSIT and $\epsilon = 1$ means the CSIT is completely unreliable.}
\label{BERK3NT2NR2}
\end{figure}

The solid lines in Fig.~\ref{BERK3NT2NR2} are the BERs of a 3-user $2
\times 2$ interference network with perfect CSIT. The analytical BER
expression of IA with MinIL agrees very well with the simulation. The
diversity order of MinIL, Max-SINR and SM are all one, which verifies
the analyses in Section \ref{SecBER}.

As shown in Fig.~\ref{BERK3NT2NR2}, IA with MinIL outperforms SM with
SVD with about 2 dB SNR gain. This is because a smaller modulation
size is used by MinIL while achieving the same data rate. In
particular, MinIL uses 4-QAM in each user pair, while SM uses 8-QAM on
every spatial channel. Moreover, even if both modes use the same modulation, 4-QAM, recalling (\ref{BERIAQPSKAPP}) and
(\ref{BERSVDQPSKAPP}) in Section \ref{SecBER}, the effective power
gain ratio of MinIL and SM is approximately $\frac{N^2}{K} =
\frac{4}{3}$. This power gain also reduces the BER of MinIL
relative to SM.

Compared with the above two modes, Max-SINR has much lower BER. For
instance, given BER around $10^{-2}$, Max-SINR has more than 7 dB SNR
gain over MinIL. This is because Max-SINR makes more effective use of
the desired channel, giving rise to the substantial coherent combining
gain. Therefore, although MinIL and Max-SINR achieve the same
multiplexing gain, Max-SINR is much superior than MinIL in terms of
reliability.

\begin{figure}
\centering
\includegraphics[width=3.7in]{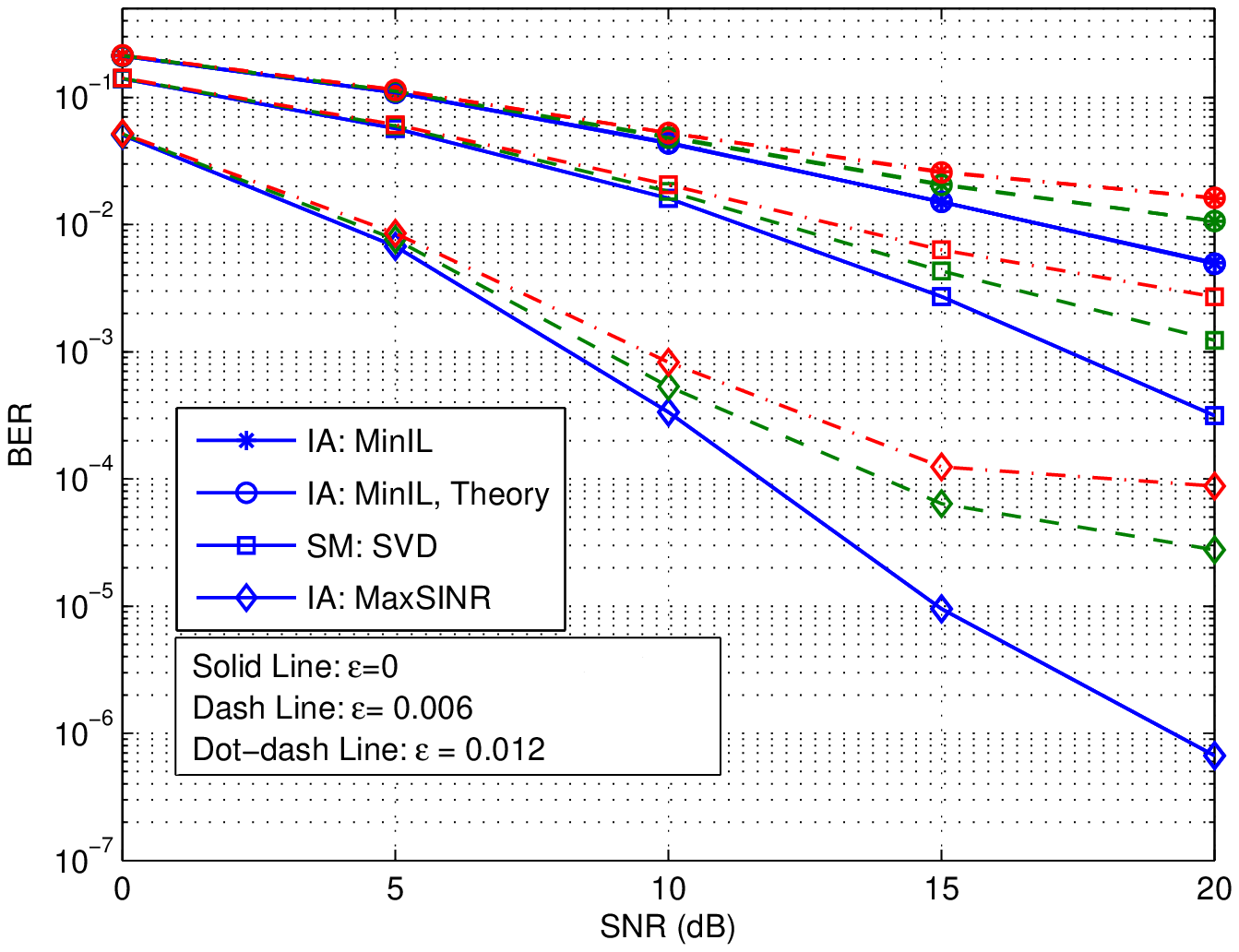}
\caption{BER performances of the three transmission modes without bit loading: $K = 3, N_t = 3, N_r =2$.}
\label{BERK3NT3NR2}
\end{figure}

Next, we consider a 3-user asymmetric $3 \times 2$ interference
network. From Fig.~\ref{BERK3NT3NR2}, both Max-SINR and SM achieve
diversity gain, while the diversity order of MinIL is still limited to
one. Since in such a system configuration, $N_t + N_r = K +2$,
Max-SINR extracts diversity gain when operating at multiplexing gain
of three~\cite{IADiverisity2011}. Max-SINR attains the best BER
performance among the three, showing around 9 dB SNR gain relative to
SM. For SVD-based SM, diversity order $(N_{\mathrm{max}}
-N_{\mathrm{min}} + 1)= 2$ is achieved. Due to the diversity gain as
well as coherent combining gain produced by an extra transmit antenna
\cite{BERSVD2006}, SM achieves smaller BER than MinIL for all considered SNRs.

\begin{figure}
\centering
\includegraphics[width=3.7in]{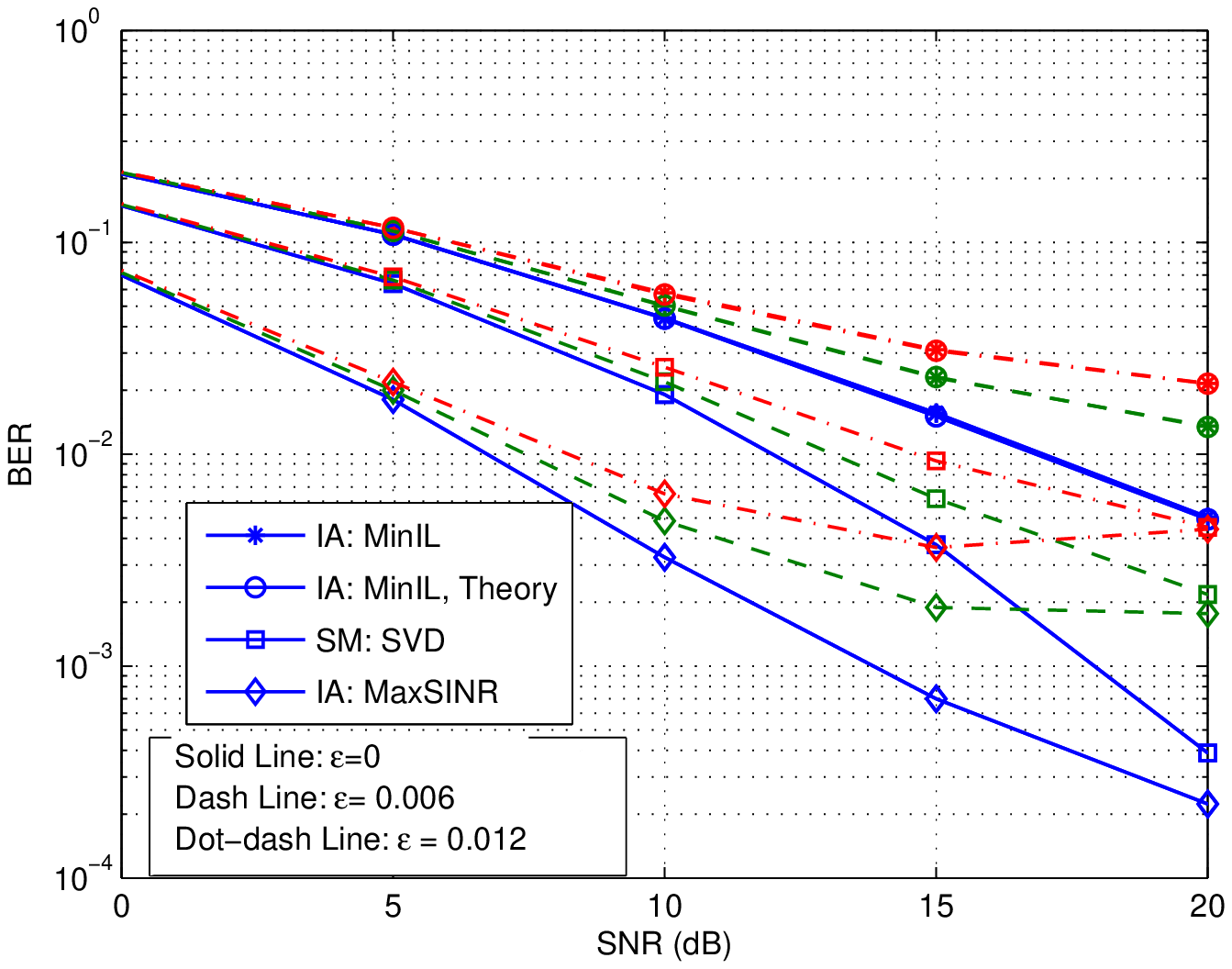}
\caption{BER performances of the three transmission modes without bit loading: $K = 4, N_t =3, N_r =2$.}
\label{BERK4NT3NR2}
\end{figure}

Note that with a $3 \times 2$ MIMO configuration, IA is capable to
accommodate 4 user pairs based on the feasibility condition
\cite{YJafar2010}. With the full multiplexing gain being exploited,
the diversity order of IA with either MinIL or Max-SINR is
one. However, in this case SM has diversity order
two. Fig.~\ref{BERK4NT3NR2} shows that although SM has a steeper slope
of BER curve (higher diversity order), Max-SINR achieves lower BER
than SM for low and moderate SNR. Intuitively, this is because
Max-SINR has large multiplexing gain as well as substantial coherent
combining gain.

\begin{figure}
\centering
\includegraphics[width=3.7in]{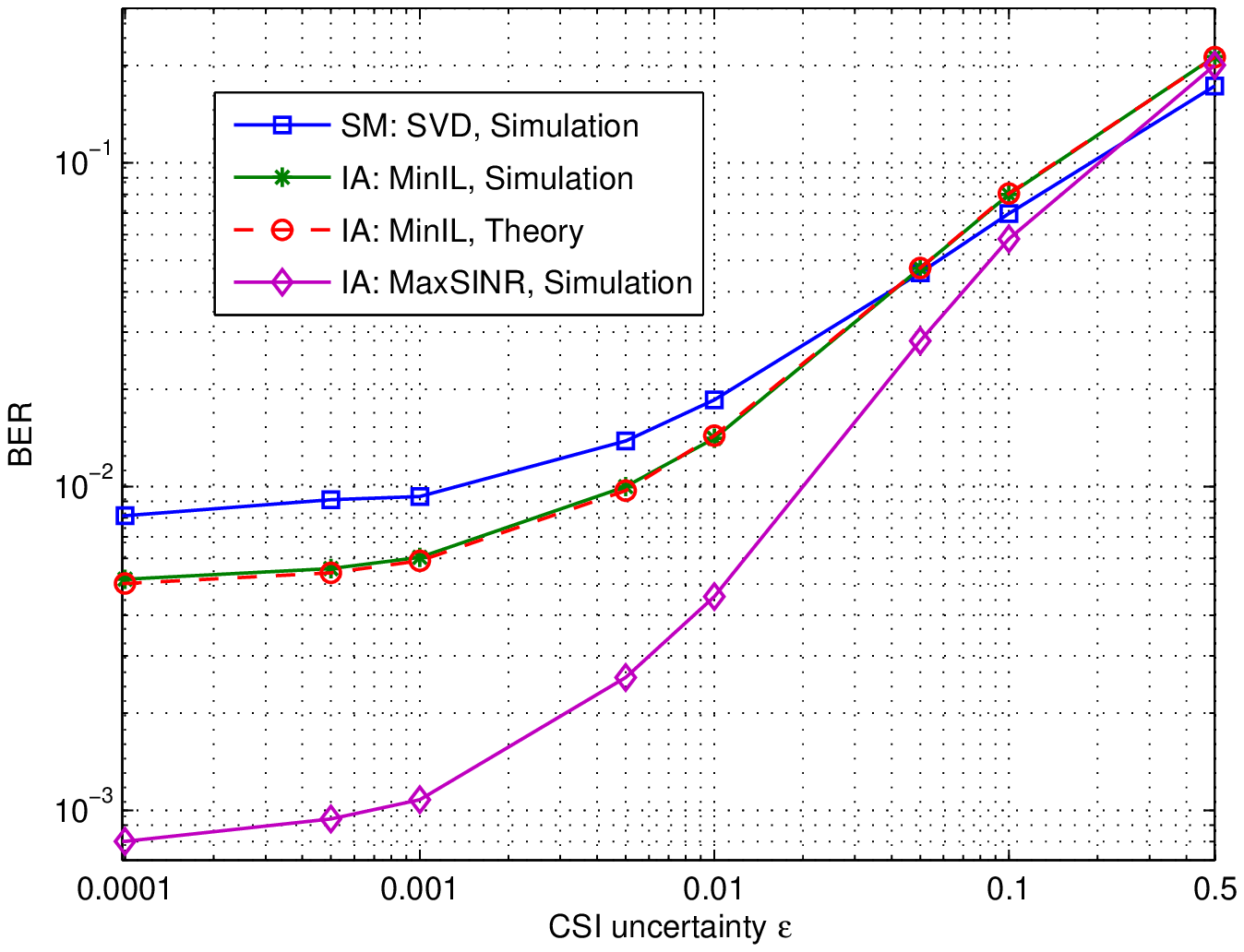}
\caption{BER performances of the three transmission modes in the presence of CSI uncertainty: $K = 3, N_t = N_r =2, \mathrm{SNR} = 20$ dB}
\label{BERK3NT2NR2VaryE}
\end{figure}

\subsubsection{CSI Uncertainty}
Now, we investigate the effects of imperfect CSIT. Based on the dash
and dot-dash lines in Figs.~\ref{BERK3NT2NR2} - \ref{BERK4NT3NR2},
with imperfect CSIT, all three modes have error floors at high SNR. As
the CSI uncertainty increases, the SNR where the error floor starts to
occur is reduced. To provide an alternative view,
Fig.~\ref{BERK3NT2NR2VaryE} presents the BER curves versus difference
levels of CSI uncertainty for a symmetric 3-user $2 \times 2$
interference network at SNR of 20 dB. The BER performances of the
three modes all degrade as CSI uncertainty grows. Among the three,
Max-SINR is most sensitive to CSI uncertainty, while SM is the most
robust one. However, Max-SINR still outperforms the other two modes
for $\epsilon \leq 0.1$. When $\epsilon >0.1$, all the three modes
have almost the same level of poor BER.



To sum up, given the same data rate and sum power, MinIL outperforms
SM for symmetric MIMO channels, while SM is better than MinIL for
asymmetric MIMO channels. Max-SINR attains the smallest BER for low to
intermediate SNR, regardless of channel configuration, but it is most
vulnerable to CSI uncertainty.

 \subsection{Performances of IA and SM with Bit Loading}\label{SecSimuBL}
 \subsubsection{Perfect CSIT}
\begin{figure}
\centering
\includegraphics[width=3.7in]{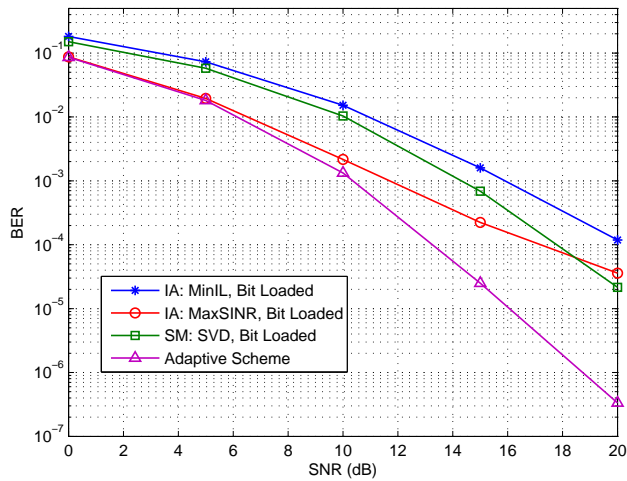}
\caption{BER performances of the three transmission modes with bit loading and the adaptive transmission scheme for perfect CSIT: $K = 3, N_t = N_r =2$}
\label{BERK3NT2NR2_BL_E0}
\end{figure}

From Fig.~\ref{BERK3NT2NR2_BL_E0} where bit loading is applied, BER is
significantly reduced compared with no bit loading
(Fig.~\ref{BERK3NT2NR2}). For example, with perfect CSIT, for SNR at
15 dB, the BERs of MinIL, Max-SINR, and SVD-based SM decrease to $2
\times 10^{-3}$, $2 \times 10^{-4}$, and $7 \times 10^{-4}$, from $1.5
\times 10^{-2}$, $2.5 \times 10^{-3}$, and $2.5 \times 10^{-2}$,
respectively. In fact, bit loading improves the error performances of
MinIL and Max-SINR by 6 dB and 4 dB SNR gain, respectively, given an
BER of $10^{-2}$.  Moreover, the diversity orders of all three modes
go beyond one. Diversity gains are achieved for all three modes: the
bit loading algorithm allows the transmitter to transmit along only
one equivalent channel if others are under deep fades. After bit
loading, SM outperforms MinIL, even for a symmetric MIMO channel. To
see this, consider the case when only one equivalent channel is
activated for both MinIL and SM. Now, IA reduces to point-to-point
MIMO channel with a single equivalent Rayleigh fading channel
$\mathbf{u}_{k}^{\dag}\mathbf{{H}}_{kk}\mathbf{v}_{k}$. In contrast,
SM transmits at the maximum eigenmode, achieving a larger diversity
gain and power gain~\cite{SVDDiversity2006}.

For Max-SINR with bit loading, its BER remains the smallest among
all three schemes for low to intermediate SNR, but is inferior to SM
for high SNR. This is because after bit loading the diversity order of
SM is about $N_tN_r = 4$, which is larger than that of Max-SINR. The
interference alignment constraints reduce the capability of Max-SINR to explore
the diversity.

The adaptive transmission scheme attains the lowest BER among all the
modes. For example, in Fig.~\ref{BERK3NT2NR2_BL_E0}, about 5 dB SNR
gain is achieved for BER at $2 \times 10^{-5}$. This additional gain
of the adaptive scheme comes from the better exploitation of the
available CSI as well as the selection of the best transmission mode
for each channel realization.

\subsubsection{CSI Uncertainty}
\begin{figure}
\centering
\includegraphics[width=3.7in]{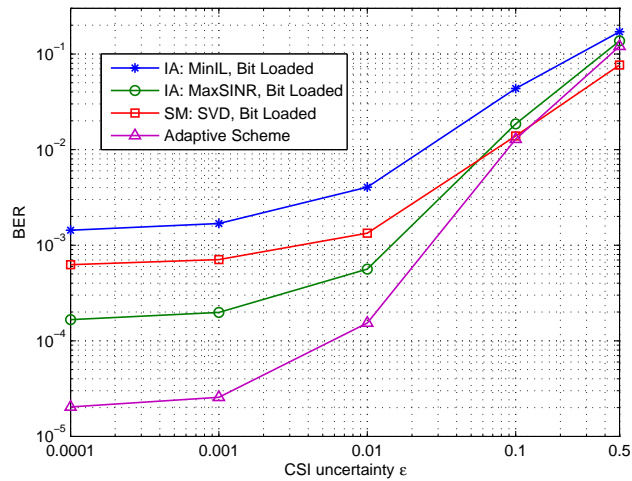}
\caption{BER performances of the three transmission modes with bit loading and the adaptive transmission scheme in the presence of CSI uncertainty: $K = 3, N_t = N_r =2, \mathrm{SNR} = 15$ dB}
\label{BERK3NT2NR2_BL_VaryE}
\end{figure}

In Fig.~\ref{BERK3NT2NR2_BL_VaryE}, we illustrate the BER performance
for bit loaded cases when there is CSI uncertainty. Here, we consider
a 3-user $2 \times 2$ interference network with SNR 15 dB. Since the
bit loading algorithms are all based on the imperfect CSIT, all the
three modes and the adaptive transmission scheme degrade as the CSI
uncertainty grows. It is interesting to note that the adaptive
transmission scheme not only reduces the error rate but also enhances
robustness to CSI uncertainty compared with Max-SINR. In this example,
the adaptive transmission scheme is better than SM when $\epsilon <
0.15$, while Max-SINR outperforms SM only when $\epsilon \leq 0.05$.

\vspace{-0.2in}

\section{Conclusion}\label{SecCon}
This paper investigates the BER performances of IA schemes and the
impact of CSI uncertainty, and in addition proposes bit loading
algorithms for IA as well as an adaptive transmission scheme. Two
representative IA algorithms, MinIL and Max-SINR, are studied. We
compare the BER performances of the two with another transmission
mode, SM with SVD. Max-SINR always outperforms the other two for low to intermediate SNR
but it is
sensitive to CSI uncertainty. Specifically, Max-SINR is superior to
other schemes as long as the CSI uncertainty is less than 10\%.  If
the CSI uncertainty is above 10\%, all three schemes have
approximately the same (poor) BER performance. Our proposed IA
bit-loading algorithm significantly improves the error performances of
MinIL and Max-SINR. Adaptive transmission achieves an even better
performance than the best of the three individual transmission modes
(with bit loading) and offers robustness to CSI uncertainty as well.

\bibliographystyle{IEEEtran}
\bibliography{IEEEabrv,IA}


\end{document}